\documentclass[aps,reprint, onecolumn, superscriptaddress,floatfix,showpacs ,pra]{revtex4-2}
\usepackage{graphicx}
\usepackage[hidelinks, colorlinks]{hyperref}
\hypersetup{citecolor=blue}
\usepackage{amsfonts, amsmath, amsthm, amssymb, bm}
\usepackage{orcidlink}
\usepackage{amsthm}

\usepackage{mathtools}
\mathtoolsset{showonlyrefs}

\usepackage{derivative}

\newcommand{\calculate}[1]{}
\usepackage{mathtools}
\usepackage[dvipsnames]{xcolor}

\newcommand{\trace}[1]{\mathrm{tr}\left({#1}\right)}
\newcommand{\ket}[1]{\left|{#1}\right\rangle}

\newcommand{\expval}[1]{\left\langle{#1}\right\rangle}

\newcommand{\comm}[2]{\left[{#1},{#2}\right]}

\newcommand{\aop}{\hat{a}}

\newcommand{\adop}{\aop^\dagger}

\newcommand{\dStar}{\mathbin{\widehat{\star}_s}}
\newcommand{\aopadop}{\left(\aop,\adop\right)}
\newcommand{\alphacoord}{\left(\alpha,\alpha^*\right)}

\newcommand{\Star}{\mathbin{\star_s}}

\newcommand{\Rdiff}[1]{\mathop{\overset{\rightarrow}{\partial}_{#1}}}
\newcommand{\Ldiff}[1]{\mathop{\overset{\leftarrow}{\partial}_{#1}}}

\newcommand{\PT}[1]{\left({#1}\right)}
\newcommand{\SB}[1]{\left[{#1}\right]}
\newcommand{\CB}[1]{\left\{{#1}\right\}}

\newcommand{\Fop}{\hat{F}}
\newcommand{\Gop}{\hat{G}}

\begin{document}

\title{Canonical quantization of a product within the Cahill-Glauber correspondence}

\author{Hendry M. Lim\orcidlink{0000-0002-2957-2438}}
\affiliation{Research Center for Quantum Physics, National Research and Innovation Agency (BRIN), South Tangerang 15314, Indonesia}
\email{hendry.minfui.lim@u.nus.edu}
\altaffiliation[Present address:~]{Department of Materials Science and Engineering, National University of Singapore, Singapore 117575, Republic of Singapore; and Center for Quantum Technologies, National University of Singapore, Singapore 117543, Republic of Singapore.}

\begin{abstract}
     The (canonical) quantization of a product of two phase-space functions can be expressed as a commutative mapping between the quantizations of its factors. In this note, we discuss such a mapping within the Cahill-Glauber $s$-parameterized correspondence framework, providing a heuristic (hence less mathematically rigorous, but arguably more easily understood) derivation for the bidifferential form. The term ``hatted star product'' is coined by virtue of the structural similarity to the star product in phase space, a special case of which is the celebrated Moyal star product.
\end{abstract}

\maketitle

Introduced by Dirac~\cite{Dirac1925, dirac1981principles}, canonical quantization (referred to as ``quantization'' herein) strives to ``make classical mechanics quantum'' by formulating a mapping between phase space functions and operators acting on the Hilbert space of quantum states, preserving the algebra of observables encoded in the Poisson bracket. That is, for an unipartite system, we look for a mapping $\mathcal{Q}$ such that
\begin{equation} \label{eq:quantization_of_PB}
    \mathcal{Q}\left(\left\{f,g\right\}_\mathrm{PB}\right) = \frac{1}{i\hbar}\left[ \mathcal{Q}\left(f\right), \mathcal{Q}\left(g\right) \right],
\end{equation}
where $\left\{f,g\right\}_\mathrm{PB} = \partial_q f \partial_p g - \partial_p f \partial_q g$ is the Poisson bracket and $\left[\hat{A},\hat{B}\right]=\hat{A}\hat{B}-\hat{B}\hat{A}$ is the commutator with $\left[\hat{q},\hat{p}\right]=i\hbar$, given the canonical position $q$ and momentum $p$. One obvious confusion from mapping scalars to operators is the ordering ambiguity: it is equally valid to map $qp$ onto $\hat{q}\hat{p}$ or $\hat{p}\hat{q}$ or $0.7\hat{q}\hat{p} + 0.3\hat{p}\hat{q}$, for instance. We may thus further ask whether any ordering choice can fulfill the requirement above. Unfortunately, it was shown by Groenewold~\cite{Groenewold1946} and van Hove~\cite{VanHove} that such a map does not globally exist beyond functions quadratic in $(q,p)$, for any operator ordering.

In addition to satisfying Eq.~\eqref{eq:quantization_of_PB}, the quantization of quadratic functions with different operator orderings gives the same result so far as commutators are concerned, hence producing identical physics. This successful quantization is reflected in the quantum harmonic oscillator having properties that incorporate those of its classical counterpart, such as the expectation value of the phase point $\left(\expval{x},\expval{p}\right)$ evolving identically to the classical phase point~\cite{curtright2013concise}. 

Although the quantization of higher-order polynomials is obstructed from satisfying Eq.~\eqref{eq:quantization_of_PB}, it may still produce interesting physics that depends on the operator ordering choice, making it worthwhile. The ordering dependence may be interpreted as there being multiple quantum systems which produce identical behavior in the classical regime. Indeed, there is no reason for a unique quantization map to exist if we view classical mechanics as some limiting behavior of quantum systems. 

Furthermore, it was revealed that quantization forms a correspondence with the phase space representation of quantum systems, one valid formulation of quantum mechanics~\cite{styer_nine_2002} originally developed by Wigner to furnish the phase space with quantum features~\cite{Wigner1932}. That is, given the quantization map $\mathcal{Q}$ tied to an operator ordering choice, relaxed from Eq.~\eqref{eq:quantization_of_PB}, its inverse $\mathcal{Q}^{-1}$ serves to transform quantum descriptions onto the phase space. 

Arguably, these interpretations are not fair as is: how can quantization make a classical observable quantum while its inverse is merely a change in representation? From a purely mathematical perspective, the correspondence is simply one established between phase and Hilbert space, such that a phase space monomial is mapped to its operator substitution under the associated ordering. The physical significance of these mappings lies in the context: both classical and quantum theories can be built upon their own sets of axioms (i.e., postulates), thereby allowing us to speak of ``phase-space-native'' and ``Hilbert-space-native'' functions of observables, i.e., those coming from the axioms. If we interpret transforming a Hilbert-space native onto the phase space as obtaining its phase space representation, then quantization, transforming a phase-space native onto the Hilbert space, can be thought of as representing it in the Hilbert space (see also the argument of Bracken~\cite{A.J.Bracken2003}). Meanwhile, a Hilbert space expression obtained from $\mathcal{Q}$ is not native in this sense, so $\mathcal{Q}^{-1}$ serves to ``dequantize'' that expression. This notion applies similarly to a phase-space function that is obtained from $\mathcal{Q}^{-1}$, for which $\mathcal{Q}$ simply returns the Hilbert-space native from which the function comes. Each mapping plays two physically different roles depending on the input, and quantization can be thought of as a representation rather than a promotion.

Setting interpretations aside, the focus of this text is to use the correspondence to study the product rule for quantization. The $\mathcal{Q}^{-1}$ mapping of an ordinary product $\hat{F}\hat{G}$ of functions in $\PT{\hat{q},\hat{p}}$ can be expressed as a noncommutative mapping between the $\mathcal{Q}^{-1}$ mappings of its factors $\hat{F}$ and $\hat{G}$. For example, in the Wigner representation~\cite{Wigner1932}, which corresponds to the Weyl quantization~\cite{Weyl1927} associated with the symmetric ordering (i.e., $qp\mapsto \PT{\hat{q}\hat{p}+\hat{p}\hat{q}}/2$), such mapping was found to be the celebrated Moyal star product~\cite{Moyal1949}, $    \mathcal{Q}^{-1}_\mathrm{W}\PT{\Fop\Gop} = \mathcal{Q}^{-1}_\mathrm{W}\PT{\Fop} \star \mathcal{Q}^{-1}_\mathrm{W}\PT{\Gop}$. In the context of the correspondence, it is then natural to ask: what is the composition rule for $\mathcal{Q}_\mathrm{W}$? The Weyl quantization of a product of scalar functions $f$ and $g$ can be expressed as a commutative, associative mapping ${\odot}$ such that $\mathcal{Q}_\mathrm{W}\PT{fg} = \mathcal{Q}_\mathrm{W}\PT{f} \mathbin{\odot} \mathcal{Q}_\mathrm{W}\PT{g}$. The study of this mapping can be traced to as early as Ref.~\cite{Omori1991}, with notable developments by Refs.~\cite{vercin1999_diamondprod, Omori2002-hg, A.J.Bracken2003, Dubin2004}, to name a few. In particular, like the Moyal star product, $\odot$ also admits a bidifferential form that applies within a specific domain, as written out in Ref.~\cite{Omori2002-hg}. More recent works, such as Refs.~\cite{Soloviev2015, bergeron2017_sStarProd}, have expanded $\star$ and $\odot$ to the more general Cahill-Glauber correspondence~\cite{Cahill1969, Cahill1969.II}, which includes the symmetric ordering. It is this correspondence that we shall discuss in this work. 

Let us define the ladder operators
\begin{equation}
    \aop = \frac{\lambda\hat{q} + i\lambda^{-1}\hat{p}}{\sqrt{2\hbar}}, \quad \adop = \frac{\lambda\hat{q} - i\lambda^{-1}\hat{p}}{\sqrt{2\hbar}},
\end{equation}
where $\lambda \in \mathbb{R}_+$, satisfying $\comm{\aop}{\adop}=1$. Let $\alpha$ and $\alpha^*$ be their scalar counterparts, serving as the variables in the $\alphacoord$ complexified phase space in which they are formally treated as independent, such that, for instance, $\partial_\alpha \alpha^*=0$ and $\partial_{\alpha^*}\SB{\PT{\alpha^*}^2\alpha} =2\alpha^*\alpha$. This phase space is more convenient to work with, and is only a linear transformation away from the canonical $\PT{q,p}$ phase space.

The Cahill-Glauber (CG) correspondence~\cite{Cahill1969, Cahill1969.II} is a family of $\mathcal{Q}$ mappings and their inverses that incorporates the normal ($\alpha^*\alpha \mapsto \adop\aop$), symmetric ($\alpha^*\alpha \mapsto \PT{\aop\adop+\adop\aop}/2$), and antinormal ($\alpha^*\alpha \mapsto \aop\adop$) ordering choices. While the formal definition is analytic, it suffices that we consider these properties:
\begin{itemize}
    \item The mapping $\mathcal{Q}_s$ is designed to map the monomial $\PT{\alpha^*}^m\alpha^n$ to its $s$-ordered operator substitution,
    \begin{equation}
        \CB{\PT{\adop}^m\aop^n}_s = \sum_{k=0}^{\min\PT{m,n}} k! \binom{m}{k} \binom{n}{k} \PT{\frac{1-s}{2}}^k \PT{\adop}^{m-k}\aop^{n-k},
    \end{equation}
    where $s\in\SB{-1,1}$ is called the order parameter. The cases $s=-1,0,1$ correspond to antinormal, symmetric, and normal ordering choices, respectively. In particular, $\mathcal{Q}_0$ denotes the Weyl quantization.
    \item Under $\mathcal{Q}^{-1}_s$, the density operator $\rho$ is mapped to a quasiprobability distribution (one that violates at least one of Kolmogorov's probability axioms) $W_{-s}\PT{\alpha,\alpha^*}$, which is the phase space representation of a quantum state. The case $s=-1$ corresponds to $\pi$ times the Glauber-Sudarshan $P$~\cite{Glauber1963, Sudarshan1963}, while the cases $s=0,1$ correspond to the Wigner $W$~\cite{Wigner1932} and Husimi $Q$~\cite{Husimi1940} representations, respectively~\footnote{%
    In their works~\cite{Cahill1969, Cahill1969.II}, Cahill and Glauber did not flip the sign of $s$ to $-s$ for the subscript of $W$. Here, we do the flip because the density operator is transformed differently from a function $\Fop\aopadop$ in the ladder operators, having a transform kernel with a flipped sign for $s$. In this work, we flip the sign for $s$ on the quasiprobability distributions to reflect our choice of having $\rho$ being mapped under $\mathcal{Q}^{-1}_s$ like $\Fop\aopadop$. The sign flip is relevant for the optical equivalence theorem: let $\expval{\Fop}=\trace{\rho \Fop}$ be the expectation value for the quantity represented by $\Fop$. Then,
    \begin{equation}
        \expval{\CB{\PT{\adop}^m\aop^n}_s} = \int_{\mathbb{R}^2} \frac{\odif[order=2]{\alpha}}{\pi} \PT{\alpha^*}^m \alpha^n W_s\alphacoord . 
    \end{equation}
    That is, while $W_s$ is mapped the same way as $\PT{-s}$-ordered operators, it serves as the weight function to calculate the expectation value of $s$-ordered operators in terms of their $\mathcal{Q}_s^{-1}$ mapping.
    }%
    . 
    \item Under $\mathcal{Q}_s$, the derivatives $\partial_\alpha f$ and $\partial_{\alpha^*} f$ are mapped onto the commutators $\comm{\mathcal{Q}_s\PT{f}}{\adop} \equiv \partial_{\aop}\mathcal{Q}_s\PT{f}$ and $\comm{\aop}{\mathcal{Q}_s\PT{f}}\equiv \partial_{\adop}\mathcal{Q}_s\PT{f}$, respectively. Here, we have used the notation for symbolic differentiation with respect to the ladder operators, applying to any power series in $\aopadop$.
    \item The mapping $\mathcal{Q}_s$ and its inverse $\mathcal{Q}^{-1}_s$ are complex-linear.
\end{itemize}
The condition for the existence of the correspondence generally depends on both $s$ and the function in question. In particular, looking from the Hilbert-space side, $\mathcal{Q}_s$ and $\mathcal{Q}_s^{-1}$ are well-behaved for any $s$-ordered power series $\sum_{m,n} c_{mn}\CB{\PT{\adop}^m\aop^n}_s$ with $c_{mn}\in\mathbb{C}$. Given two such functions $\Fop\aopadop$ and $\Gop\aopadop$, Soloviev~\cite{Soloviev2015} showed that
\begin{equation}
    \mathcal{Q}^{-1}_s \PT{\Fop\Gop} = \mathcal{Q}^{-1}_s \PT{\Fop} \Star \mathcal{Q}^{-1}_s\PT{\Gop},
\end{equation}
where the $s$-parameterized star product takes the differential form
\begin{equation}\label{eq:star_product}
    \Star = \exp\PT{
    \frac{s+1}{2}\Ldiff{\alpha} \Rdiff{\alpha^*}
    +
    \frac{s-1}{2}\Ldiff{\alpha^*} \Rdiff{\alpha}
    }.
\end{equation}
The case $s=0$ is the Moyal star product $\star$ we discussed above. The derivative operators with overset arrows are understood as doing derivatives with respect to the variable belonging to the expression positioned to the left or right of $\Star$. To see this property more intuitively, we may consider the ``Bopp shift''~\cite{Hillery1984Distribution, curtright2013concise} evaluation of the star product, based on the property that $\exp\PT{\alpha\partial_x}f(x)=f(x+\alpha)$ with $\alpha$ independent of $x$. Given two phase space functions $f\alphacoord$ and $g\alphacoord$, we have
\begin{equation}
\begin{split}
    f\alphacoord \Star g\alphacoord 
    &= 
    f\PT{\alpha + \frac{s+1}{2}\Rdiff{\alpha^*}, \alpha^* + \frac{s-1}{2}\Rdiff{\alpha}} g\alphacoord
    \\&=
    f\PT{\alpha + \frac{s+1}{2}\partial_{\beta^*}, \alpha^* + \frac{s-1}{2}\partial_{\beta}} g\PT{\beta,\beta^*} \Bigg|_{\beta\mapsto\alpha, \beta^*\mapsto\alpha^*}
    \\&= 
    f\alphacoord g\PT{\alpha + \frac{s-1}{2}\Ldiff{\alpha^*}, \alpha^* + \frac{s+1}{2}\Ldiff{\alpha}}
    \\&=
    g\PT{\alpha + \frac{s-1}{2}\partial_{\beta^*}, \alpha^* + \frac{s+1}{2}\partial_{\beta}} f\PT{\beta,\beta^*} \Bigg|_{\beta\mapsto\alpha, \beta^*\mapsto\alpha^*}.
\end{split}
\end{equation}
As an example, it is well-known that the expression $\adop\aop\rho$ corresponds to $\PT{\alpha^*-\partial_\alpha}\PT{\alpha \pi P}$, when we take $\pi P$ as the weight function~\footnote{The $1/\pi$ factor in the expansion integral cancels the $\pi$ of the weight function, hence the typical expression without $\pi$ anywhere in the integral.} for expanding the density operator in terms of the coherent states $\ket{\alpha}$, which are the eigenstates of $\aop$~\cite{gardiner_2004_QuantumNoiseHandbook}. We can arrive at this correspondence within the CG framework by setting $s=-1$, in which case $\adop\aop=\CB{\adop\aop}_{-1}-1$ is mapped to $\alpha^*\alpha-1$, while $\rho$ is mapped to $\pi P$. It is then straightforward to verify that $\PT{\alpha^*\alpha-1}\mathbin{\star_{-1}} \PT{\pi P} = \PT{\alpha^*-\partial_\alpha}\PT{\alpha \pi P}$. 

With the star product in mind, we are ready to heuristically derive the differential form of the generalization of the $\odot$ mapping within the CG correspondence. Let us denote this mapping by $\dStar$, and call it the ``hatted star product'', for a reason that will become apparent below. Given two phase-space functions $f\alphacoord$ and $g\alphacoord$ that admit power series expansions in $\alphacoord$, we define the mapping $\dStar$ by
\begin{equation}
\begin{split}
    \mathcal{Q}_s\PT{fg} &= \mathcal{Q}_s\PT{f} \dStar \mathcal{Q}_s\PT{g} \\&= \mathcal{Q}_s\PT{g} \dStar \mathcal{Q}_s\PT{f}.
\end{split}
\end{equation}
Like the star product $\Star$, the mapping $\dStar$ is complex-bilinear, 
\begin{equation}
\begin{split}
    \SB{c_1\mathcal{Q}_s\PT{f_1} + c_2\mathcal{Q}_s\PT{f_2}} \dStar \mathcal{Q}_s\PT{g}
    &= \mathcal{Q}_s \PT{c_1f_1+c_2f_2} \dStar \mathcal{Q}_s\PT{g}
    \\
    &= \mathcal{Q}_s \PT{\SB{c_1f_1+c_2f_2}g}
    \\
    &=
    c_1 \mathcal{Q}_s\PT{f_1 g}+c_2\mathcal{Q}_s\PT{f_2g}
    \\&=
    c_1 \SB{\mathcal{Q}_s\PT{f_1}\dStar \mathcal{Q}_s\PT{g}} + c_2 \SB{\mathcal{Q}_s\PT{f_2}\dStar \mathcal{Q}_s\PT{g}},
\end{split}
\end{equation}
for $c_j\in \mathbb{C}$, and, more importantly, associative,
\begin{equation}
\begin{split}
    \mathcal{Q}_s\PT{fgh} &=
    \mathcal{Q}_s\PT{fg}\dStar \mathcal{Q}_s\PT{h} 
    \\ &= 
    \SB{\mathcal{Q}_s\PT{f}\dStar \mathcal{Q}_s\PT{g}} \dStar \mathcal{Q}_s\PT{h}
    \\&=
    \mathcal{Q}_s\PT{f}\dStar \SB{\mathcal{Q}_s\PT{g} \dStar \mathcal{Q}_s\PT{h}}
    \\&=
    \mathcal{Q}_s\PT{f}\dStar \mathcal{Q}_s\PT{gh}.
\end{split}
\end{equation}
Based on these properties, it suffices that we find an explicit form of $\dStar$ which satisfies
\begin{equation}\label{eq:dStar_condition_1}
    \aop\dStar \mathcal{Q}_s\PT{g} = \mathcal{Q}_s\PT{g} \dStar \aop = \mathcal{Q}_s\PT{\alpha g},
\end{equation}
and 
\begin{equation}\label{eq:dStar_condition_2}
    \adop\dStar \mathcal{Q}_s\PT{g} = \mathcal{Q}_s\PT{g} \dStar \adop = \mathcal{Q}_s\PT{\alpha^* g},
\end{equation}
since we can always write
\begin{equation}
\begin{split}
    \mathcal{Q}_s\PT{\sum_{m,n} c_{mn} \SB{\alpha^*}^m\alpha^n f}
    &=
    \sum_{m,n} \SB{c_{mn} \mathcal{Q}_s\PT{\alpha} \dStar \mathcal{Q}_s\PT{\SB{\alpha^*}^m\alpha^{n-1} f}}
    \\&=
    \sum_{m,n} \SB{c_{mn}\aop \dStar \mathcal{Q}_s\PT{g_{mn}}}
\end{split}
\end{equation}
for $n\geq 1$, and likewise for $\adop$. Put another way, different ``hatted-star-factorizations'' of a given $\mathcal{Q}_s$ mapping are equivalent to each other; if $\dStar$ holds for one, then it must hold for the others. To come up with an explicit form for $\dStar$, let us consider the notation $\Fop=\mathcal{Q}_s\PT{f}$ to avoid clutter and write 
\begin{equation}
\begin{split}
    \mathcal{Q}_s^{-1}\PT{\aop\Fop} 
    &= 
    \alpha \Star \mathcal{Q}_s^{-1}\PT{\Fop} 
    \\&= 
    \alpha f +\frac{s+1}{2} \partial_{\alpha^*} f.
\end{split}
\end{equation}
Rearranging gives
\begin{equation}
    \mathcal{Q}_s^{-1}\PT{\aop\Fop} - \frac{s+1}{2}\partial_{\alpha^*} f = \alpha f.
\end{equation}
Using the property $\mathcal{Q}_s\PT{\partial_{\alpha^*}f}=\partial_{\adop} \Fop$ discussed above, we obtain
\begin{equation}
    \mathcal{Q}_s^{-1}\PT{\aop\Fop - \frac{s+1}{2}\partial_{\adop}\Fop} = \alpha f.
\end{equation}
In order for Eq.~\eqref{eq:dStar_condition_1} to hold, we must have
\begin{equation}
    \aop \dStar \Fop = \PT{\aop - \frac{s+1}{2} \partial_{\adop}} \Fop.
\end{equation}
Following similar steps, it can be shown that
\begin{align}
     \Fop \dStar \aop &= \PT{\aop-\frac{s-1}{2}\partial_{\adop}} \Fop,
    \\
    \adop\dStar \Fop &= \PT{\adop - \frac{s-1}{2}\partial_{\aop}} \Fop, 
    \\
    \Fop \dStar \adop &= \PT{\adop - \frac{s+1}{2} \partial_{\aop}} \Fop
\end{align}
for Eqs.~\eqref{eq:dStar_condition_1} and~\eqref{eq:dStar_condition_2} to hold. These equations look just like the Bopp shift evaluations of $\alpha\Star f$, $f\Star\alpha$, $\alpha^*\Star f$, and $f \Star \alpha^*$, respectively, but with $\aopadop$ instead of $\alphacoord$. Under the assumptions of complex-bilinearity and associativity, these four ``constituting relations'' in phase space can be used to infer Eq.~\eqref{eq:star_product}. In a similar spirit, we can thus infer the bidifferential form of $\dStar$,
\begin{equation} \label{eq:hatted_star_product}
    \dStar = \exp\PT{-\SB{\frac{s+1}{2}\Ldiff{\aop}\Rdiff{\adop} + \frac{s-1}{2}\Ldiff{\adop}\Rdiff{\aop}}}.
\end{equation}
The Bopp shift evaluation symbolically applies for $\aopadop$ as well, though care must be taken for the operator ordering. For example, since $\SB{\hat{A},\hat{B}\hat{C}} = \SB{\hat{A},\hat{B}}\hat{C} + \hat{B}\SB{\hat{A},\hat{C}}$, we have $\partial_{\adop}\PT{\hat{X}\Fop}=\hat{X}\partial_{\adop}\Fop$ when $\SB{\aop, \hat{X}}=0$. Evidently, evaluations using this form is guaranteed to converge for any two polynomials in $\aopadop$. As for infinite power series, care may be needed due to the possibility of divergence. This result agrees with Eqs. (16.3) and (19.1) in Ref.~\cite{bergeron2017_sStarProd}, which give Eq.~\eqref{eq:hatted_star_product} expanded as a power series. It is also straightforward to verify that $s=0$ gives an operator version of the ``inverse Moyal product formula'' in Ref.~\cite{Omori2002-hg}.

In the context of quantization, $\dStar$ encodes the operator ordering choice, constituting the product rule. For example, we know that the normal-ordered quantization of ${\alpha^*}\alpha^2$ is ${\adop}\aop^2$, which we can obtain using $\mathbin{\hat{\star}_1}$:
\begin{equation}
\begin{split}
    \mathcal{Q}_1\PT{\alpha^*\alpha^2} 
    &=
    \aop^2 \mathbin{\hat{\star}_1} \adop
    \\
    &= \PT{\aop - \Rdiff{\adop}}^2\adop 
    \\
    &= \CB{\SB{{\aop}^2 - \partial_{\hat{b}^\dagger} \aop - \aop\partial_{\hat{b}^\dagger} +\partial_{\hat{b}^\dagger}^2} \hat{b}^\dagger} \Bigg|_{\hat{b}\mapsto\aop, \hat{b}^\dagger \mapsto \adop}
    \\
    &= \CB{\aop^2\hat{b}^\dagger - 2 \aop \SB{\hat{b},\hat{b}^\dagger} + \comm{\hat{b}}{\comm{\hat{b}}{\hat{b}^\dagger}}} \Bigg|_{\hat{b}\mapsto\aop, \hat{b}^\dagger \mapsto \adop}
    \\
    &= \aop^2\adop - 2\aop.
\end{split}
\end{equation}
It is straightforward to verify that $\adop\mathbin{\hat{\star}_1}\aop^2$ gives an equivalent result.

As for the terminology, we emphasize how similar $\dStar$ in Eq.~\eqref{eq:hatted_star_product} looks compared to $\Star$ in Eq.~\eqref{eq:star_product}, as if one is the ``inverse'' of the other. Since $\dStar$ operates on quantum operators, which are usually written using hats, we call the mapping $\dStar$ the ``hatted star product'': a mapping between Hilbert-space operators that serves as the product rule for $\mathcal{Q}_s$, reminiscent of its phase-space counterpart that serves as the product rule for $\mathcal{Q}_s^{-1}$. 

\begin{acknowledgments}
    H.M.L. acknowledges the use of GPT-5.6 Sol, a large-language model developed by OpenAI, to rummage through the arguably obscure literature for this topic and to check for mathematical consistency throughout the discussion.
\end{acknowledgments}

\bibliography{bibliography}

\end{document}